\documentstyle[amssymb,aps]{revtex}

\draft

\begin{document}
\title{Ground State of Strongly Correlated Fermions: Short-Range Order}
\author{Yu.B.Kudasov\thanks{%
Electronic address: kudasov@ntc.vniief.ru}}
\address{Russian Federal Nuclear Center - VNIIEF, Sarov, 607190, Russia}
\maketitle

\begin{abstract}
A new variational method is developed to calculate the ground state energy
of Fermi systems with strong short-range correlations. A trial wave function
of Gutzwiller's type contains additional variational parameters
corresponding to configurations of pairs of nearest-neighbor sites. To
evaluate the ground state energy, generalized Kikuchi's pseudo-ensemble
method is used. The Hubbard model at half band-filling is investigated. The
ground state energy of the paramagnetic phase is calculated for a chain,
square and simple cubic lattices. It is shown that the short-range order
lowers drastically the ground state energy of the Hubbard model at
intermediate interaction strength. The paramagnetic phase of the
Kondo-Hubbard model ($S=1/2$ and $S=5/2$) at half band-filling is
investigated. The ground state energy, correlation functions and effective
mass are calculated for chain, square and simple cubic lattices. A phase
transition was found for simple cubic lattice.
\end{abstract}


\twocolumn

\section{Introduction}

A short-range order, i.e. strong short-range correlations, is an intrinsic
feature of strongly correlated Fermi systems. It was observed in a metal
phase of V$_{2}$O$_{3}$ - a famous Mott-Hubbard system \cite{Bao},
high-temperature superconductors \cite{HTSC}, and heavy fermion systems \cite
{Aeppli,Aeppli2,Regn,Bernh}. Principal aspects of the short-range order can
be investigated within the Hubbard model \cite{Hubbard,Gutzw,Kanam}. The
exact solution of the Hubbard Hamiltonian is known for 1D chain \cite{Lieb}.
Great simplifications appear in infinite dimensions because spatial
correlations do not play an important role in this limit \cite{Metz}. As for
lattices of intermediate dimensions, which are of great practical
importance, one has to follow various numerical and analytical
approximations \cite{Gebhard,Fulde}. Recently a new variational approach to
the ground state of the Hubbard model was proposed \cite{Kudasov}. In
addition to intrasite correlations the trial wave function of Gutzwiller's
type \cite{Gutzw} contains nearest-neighbor correlations in an explicit
form. In contrast to Ref.\cite{Fulde}, Kikuchi's (cluster variation) method 
\cite{Kik} was used to evaluate the ground state energy. It was shown that
the short-range correlations affect significantly the ground state of the
Hubbard model. A comparison of this result with the variational Monte Carlo
method (VMC) \cite{Yokoy} based on the Gutzwiller trial wave function and $%
1/D$-expansion in the dynamical mean-field theory \cite{Gebh} shows that the
latters underestimate significantly the ground state energy at intermediate
coupling.

A heavy fermion behavior usually arises from an interplay between a lattice
of localized $f$-electrons and itinerant electrons. In the Kondo limit, such
a system becomes a Kondo lattice \cite{Hewson}. In many cases the itinerant
subsystem of the Kondo lattice is formed by a narrow $d$-band where the
short-range Coulomb interaction between itinerant electrons is considerable.
There is a lot of examples of this kind \cite{Hewson}. Therefore it is
reasonable to describe the itinerant subsystem by means of the Hubbard
Hamiltonian. Thus, we come to the Kondo-Hubbard lattice model \cite{Schork}.
It was shown by means of neutron scattering experiments that strong
short-range antiferromagnetic (AFM) correlations exist in Kondo and
Kondo-Hubbard lattices \cite{Aeppli,Aeppli2,Regn,Bernh}. They play an
important role in the heavy fermions behavior. In this paper, I apply the
variational theory of Ref.\cite{Kudasov} to the Kondo-Hubbard lattice model.
In Sec.II, the variational technique is introduced. The ground state energy
of the paramagnetic (PM) phase of the Hubbard model at half-band filling is
calculated. In Sec.III, the technique is generalized to the Kondo-Hubbard
lattice model. The ground state energy, correlation functions, and effective
mass are calculated at half-band filling for (i) spin $1/2$ and (ii) spin $%
5/2$ ($f$-spin) Kondo-Hubbard lattices. The results are discussed in Sec.IV.

\section{Ground State of the Hubbard model}

\subsection{Trial Wave Function}

Let us consider a lattice with one orbital per site and restrict ourselves
to nearest-neighbor hopping only. Then, the Hubbard model has the following
form 
\begin{equation}
H_{{\normalsize H}}=t\sum_{\left\langle ij\right\rangle ,\sigma }\left(
a_{i\sigma }^{\dagger }a_{j\sigma }+\text{H.c.}\right)
+U\sum_{i}n_{i\uparrow }n_{i\downarrow }  \label{hubb}
\end{equation}
where $a_{i\sigma }^{\dagger }\left( a_{j\sigma }\right) $ is the creation
(annihilation) operator of a fermion of spin $\sigma =\uparrow ,\downarrow $
on the $i$-th lattice site, $\left\langle ij\right\rangle $ denotes a pair
of adjacent sites, $n_{i\sigma }=a_{i\sigma }^{\dagger }a_{j\sigma }$.

The Gutzwiller trial wave function \cite{Gutzw} gives a good basis for a
variational analysis of the Hubbard model ground state in infinite
dimensions. This trial function can be written in the following symbolic
form \cite{Vollhardt}

\begin{equation}
|\psi \rangle =g_{0}^{\hat{X}}|\varphi _{0}\rangle  \label{gutz}
\end{equation}
where $\hat{X}=\sum_{i}n_{i\uparrow }n_{i\downarrow }$, $g_{0}$ is the real
parameter taking a value in the interval $[0,1]$ if $U\geq 0$, $|\varphi
_{0}\rangle $ is the $N$-particle wave function of non-interacting fermions,
for instance 
\begin{equation}
|\varphi _{0}\rangle =\prod_{{\bf k}\in {\bf V}_{F\uparrow }}a_{{\bf k}%
\uparrow }^{\dagger }\prod_{{\bf k}\in {\bf V}_{F\downarrow }}a_{{\bf k}%
\downarrow }^{\dagger }|0\rangle \text{.}  \label{Fermi}
\end{equation}
Here, $a_{{\bf k}\sigma }^{\dagger }$ denotes the creation operator of the
Bloch state, ${\bf k}$ is the wave vector, and ${\bf V}_{F\sigma }$ is the
space within the Fermi surface.

To control nearest-neighbor correlations between fermions one can extend the
Gutzwiller trial wave function as \cite{Kudasov} 
\begin{equation}
\left| \psi \right\rangle =\prod_{\lambda }g_{\lambda }^{\widehat{P}%
_{\lambda }}\left| \varphi _{0}\right\rangle  \label{new}
\end{equation}
where $\widehat{P}_{\lambda }$ are the projection operators onto all
feasible configurations of a single site and a pair of nearest-neighbor
sites, $g_{\lambda }$ are the nonnegative real parameters. In the PM phase,
there are 4 such operators for intrasite configurations 
\begin{eqnarray*}
\widehat{X}_{1} &=&\sum_{i}\left( 1-n_{i\uparrow }\right) \left(
1-n_{i\downarrow }\right) , \\
\widehat{X}_{2} &=&\sum_{i}n_{i\uparrow }\left( 1-n_{i\downarrow }\right) ,
\\
\widehat{X}_{3} &=&\sum_{i}\left( 1-n_{i\uparrow }\right) n_{i\downarrow },
\\
\widehat{X}_{4} &=&\sum_{i}n_{i\uparrow }n_{i\downarrow }\text{,}
\end{eqnarray*}
and 10 operators for nearest-neighbor configurations 
\begin{eqnarray}
\widehat{Y}_{1} &=&\sum_{\left\langle ij\right\rangle }\left( 1-n_{i\uparrow
}\right) \left( 1-n_{i\downarrow }\right) \left( 1-n_{j\uparrow }\right)
\left( 1-n_{j\downarrow }\right) ,  \nonumber \\
\widehat{Y}_{2} &=&\sum_{\left\langle ij\right\rangle }n_{i\uparrow
}n_{i\downarrow }n_{j\uparrow }n_{j\downarrow }\text{, and etc.}  \nonumber
\end{eqnarray}
All the operator $\widehat{Y}_{\lambda }$ and corresponding pair
configurations are shown in Table \ref{table}.

From now on, we shall consider only lattices for which the total number of
nearest-neighbors pairs is equal to $zL/2$, where $z$ is the number of the
nearest neighbors of a site and $L$ is the total number of sites. Let us
denote normalized eigenvalues of the projection operators as $x_{\lambda
}\left| \Phi \right\rangle =L^{-1}\widehat{X}_{\lambda }\left| \Phi
\right\rangle $, $y_{\lambda }\left| \Phi \right\rangle =\left( zL/2\right)
^{-1}\widehat{Y}_{\lambda }\left| \Phi \right\rangle $. The eigenvalues turn
out to be related to each other by normalization conditions \cite{Ziman} 
\begin{equation}
\sum_{\lambda }x_{\lambda }=1\text{, }\sum_{\lambda }\beta _{\lambda
}y_{\lambda }=1  \label{norm}
\end{equation}
and self-consistency conditions \cite{Ziman} 
\begin{eqnarray}
y_{1}+y_{3}+y_{4}+y_{5} &=&x_{1}\text{,}  \nonumber \\
y_{2}+y_{3}+y_{8}+y_{9} &=&x_{4}\text{,}  \nonumber \\
y_{4}+y_{6}+y_{7}+y_{8} &=&x_{2}\text{,}  \nonumber \\
y_{5}+y_{7}+y_{9}+y_{10} &=&x_{3}\text{.}  \label{self}
\end{eqnarray}

As concentrations of fermions of each spins are fixed there are the only
independent parameter $x_{\lambda }$ and 7 independent parameters $%
y_{\lambda }$. In the case of half band-filling, additional constrains
appear 
\begin{equation}
y_{1}=y_{2}\text{, }y_{6}=y_{10}\text{, }y_{4}=y_{5}=y_{8}=y_{9}  \label{add}
\end{equation}
which reduce the number of the independent parameters $y_{\lambda }$ to 3.
Assume that $x_{1}=x_{4}=x$, $y_{3}$, $y_{4}$, and $y_{7}$ are the
independent parameters. Then, we obtain the final form of the generalized
trial wave function of the PM phase at half band-filling

\begin{equation}
\left| \psi \right\rangle =g_{0}^{\widehat{X}}g_{3}^{\beta _{3}\widehat{Y}%
_{3}}g_{4}^{4\beta _{4}\widehat{Y}_{4}}g_{7}^{\beta _{7}\widehat{Y}%
_{7}}\left| \varphi _{0}\right\rangle \text{.}  \label{CWT}
\end{equation}

To elucidate the physical meaning of the trial function (\ref{CWT}) let us
rewrite the initial wave function of non-interacting fermions as a
superposition of configurations 
\begin{eqnarray*}
|\varphi _{0}\rangle &=&\sum_{\Gamma }A_{\Gamma }|\Gamma \rangle , \\
|\Gamma \rangle &=&\prod_{i,\sigma }a_{i\sigma }^{\dagger }|0\rangle
\end{eqnarray*}
where $A_{\Gamma }$ is the complex amplitude of the configuration $\Gamma $, 
$|0\rangle $ is the vacuum state. Then, we obtain 
\begin{equation}
\left| \psi \right\rangle =\sum_{\Gamma }g_{0}^{D_{\Gamma
0}}g_{3}^{2D_{\Gamma 3}}g_{4}^{8D_{\Gamma 4}}g_{7}^{2D_{\Gamma 7}}A_{\Gamma
}\left| \Gamma \right\rangle  \label{PMWF}
\end{equation}
Here $D_{\Gamma 0}$ denotes the number of doubly occupied sites in the
configuration $\Gamma $, $D_{\Gamma 3(4,7)}$ are the numbers of
configurations of nearest-neighbor pairs corresponding to the operator $%
\widehat{Y}_{3}$, $\widehat{Y}_{4}$, and $\widehat{Y}_{7}$ in the
configuration $\Gamma $. Since other operators $\widehat{Y}_{\lambda }$ are
related to them by Eqs.(\ref{norm}), (\ref{self}), and (\ref{add}), {\em all}
the possible configurations of nearest-neighbor pairs are taken into account
explicitly.

Since the operator $\widehat{F}=g_{0}^{\widehat{X}}g_{3}^{\beta _{3}\widehat{%
Y}_{3}}g_{4}^{4\beta _{4}\widehat{Y}_{4}}g_{7}^{\beta _{7}\widehat{Y}_{7}}$
is a polynomial of $n_{i\sigma }$, it commutes with an operator of particle
alternation. Thus, the trial wave function (\ref{PMWF}) is antisymmetric. $%
\widehat{F}$ is also invariant under the operations transforming lattice
into itself, namely, translations, reflections, rotations and inversion.
This means that all the symmetries of the initial wave function remain valid
for the trial wave function.

\subsection{Ground State Energy of the PM phase}

It is well known that correlations between fermions of the same spin exist
even in the initial state (\ref{Fermi}), i.e. at $U=0$. That is why, first
we should evaluate its norm using an equiprobable state. The norm of a
correlated state is calculated also using the equiprobable state. Finally,
all expressions should be normalized to the norm of the initial state (\ref
{Fermi}). Since operators $\widehat{F\text{ }}$make up an operator group ($%
\widehat{F}(g_{0},g_{3},g_{4},g_{7})\widehat{F}(g_{0}^{\prime
},g_{3}^{\prime },g_{4}^{\prime },g_{7}^{\prime })=\widehat{F}%
(g_{0}g_{0}^{\prime },g_{3}g_{3}^{\prime },g_{4}g_{4}^{\prime
},g_{7}g_{7}^{\prime })$ and $\widehat{F}\widehat{F}^{-1}=1$ when $%
g_{\lambda }$ are finite and nonzero) this procedure is equivalent to
applying the Fermi sea state (\ref{Fermi}) as the initial one in Eq.(\ref
{CWT}). Thus, the norm of any state generated by Eq.(\ref{PMWF}) is \cite
{Kudasov} 
\begin{eqnarray}
\left\langle \psi |\psi \right\rangle &=&\sum_{\left\{
x,y_{3},y_{4},y_{7}\right\} }W_{\left\{ x,y_{3},y_{4},y_{7}\right\}
}g_{0}^{2Lx}g_{3}^{2zLy_{3}}g_{4}^{8zLy_{4}}g_{7}^{2zLy_{7}}  \nonumber \\
&=&\sum_{\left\{ x,y_{3},y_{4},y_{7}\right\} }R_{\left\{
x,y_{3},y_{4},y_{7}\right\} }\text{.}  \label{NormPM}
\end{eqnarray}
A factor which is inessential for the further calculations is dropped in Eq.(%
\ref{NormPM}). The summation is performed over all the sets $\left\{
x,y_{3},y_{4},y_{7}\right\} $. A lot of configurations are related to the
same set of the independent variables. Then, $W_{\left\{
x,y_{3},y_{4},y_{7}\right\} }$ is the number of the configurations
corresponding to the fixed set $\left\{ x,y_{3},y_{4},y_{7}\right\} $ or the
weight of this set. To calculate this quantity we use Kikuchi's
pseudo-ensemble method \cite{Kik}. It should be mentioned that this method
is practically exact for Bethe lattices and approximate for lattices with
closed paths \cite{Ziman}. According to Kikuchi's hypothesis the weight of a
set can be expressed as a product \cite{Kik}: 
\begin{equation}
W=\Gamma Q  \label{W}
\end{equation}
Here, lower indices are omitted for the sake of simplicity. The quantity $Q$
determines the number of arrangements of ten indistinguishable elements corre%
{\it s}ponding to $\widehat{Y}_{\lambda }$\ at $zL/2$\ pairs, i.e. the
multinomial coefficients 
\begin{equation}
Q=\frac{(\frac{zL}{2})!}{\prod_{\lambda }[(\frac{zy_{\lambda }L}{2}%
)!]^{\beta _{\lambda }}},  \label{Q}
\end{equation}
and 
\begin{equation}
\Gamma =\frac{L!\prod_{\lambda }(x_{\lambda }zL)!}{(zL)!\prod_{\lambda
}(x_{\lambda }L)!}  \label{gamma}
\end{equation}
is the fraction of proper arrangements in the pseudo-ensemble\cite{Ziman}.
In Eqs.(\ref{Q}) and (\ref{gamma}) the dependent variables should be
expressed in terms of $x,y_{3},y_{4},y_{7}$ as follows 
\begin{eqnarray}
x_{2} &=&x_{3}=1/2-x,  \nonumber \\
y_{1} &=&y_{2}=x-y_{3}-2y_{4},  \label{depend} \\
y_{6} &=&y_{10}=1/2-x-y_{7}-2y_{4}.  \nonumber
\end{eqnarray}
From now on, we use $y_{2}$ and $y_{6}$ as brief notations of the
expressions (\ref{depend}).

In the thermodynamic limit ($L\rightarrow \infty $) we can retain, in the
usual fashion \cite{Kik,Ziman}, only the terms of the series (\ref{NormPM})
which are very close to the largest one that is the following condition
should be valid $\{x,y_{3},y_{4},y_{7}\}\rightarrow
\{x,y_{3},y_{4},y_{7}\}_{MAX}$. All the other terms are exponentially small.
Since $R$ is a nonnegative function, it is convenient to search the global
maximum of its logarithm rather than of itself. Let us transform the
factorials involved in $R$ by means of the asymptotic Stirling formula.
Then, let us find the logarithm of $R$ and retain the leading term on $L$
only. A straightforward calculation yields 
\begin{eqnarray}
L^{-1}\ln W &=&2(z-1)[x\ln x+(1/2-x)\ln (1/2-x)]  \nonumber \\
&&-z(y_{2}\ln y_{2}+y_{3}\ln y_{3}+4y_{4}\ln y_{4}  \nonumber \\
&&+y_{6}\ln y_{6}+y_{7}\ln y_{7}).  \label{lnW}
\end{eqnarray}
The domain of function (\ref{lnW}) is limited by conditions (\ref{norm}) and
(\ref{self}). It can be shown that the gradient of the function at
boundaries is directed inwards to the domain. Therefore the global maximum
of $R$ is an internal one and conditions 
\begin{equation}
\frac{\partial (\ln R)}{\partial \eta _{\lambda }}=0\text{,}  \label{dlnR}
\end{equation}
where $\eta _{\lambda }=x,y_{3},y_{4},y_{7}$, are necessary for the global
maximum. They lead to the following system of equations that relate $g_{i}$
to $x$ and $y_{i}$%
\begin{eqnarray}
g_{0} &=&\left( \frac{1/2-x}{x}\right) ^{z-1}\left( \frac{x-y_{3}-2y_{4}}{%
1/2-x-y_{7}-2y_{4}}\right) ^{z/2},  \nonumber \\
g_{3}^{2} &=&\frac{y_{3}}{x-y_{3}-2y_{4}},  \nonumber \\
g_{4}^{4} &=&\frac{y_{4}^{2}}{\left( 1/2-x-y_{7}-2y_{4}\right) \left(
x-y_{3}-2y_{4}\right) },  \nonumber \\
g_{7}^{2} &=&\frac{y_{7}}{1/2-x-y_{7}-2y_{4}}.  \label{sys}
\end{eqnarray}

It should be mentioned that $L^{-1}\ln R$ is rigorously a convex upwards
function{\em \ }of $y_{3},y_{4},y_{7}$ at any fixed $x$. This means that, in
effect, we search the maximum of a function of the only inexplicit variable.

To calculate the ground state energy of the Hamiltonian (\ref{hubb}) we need
to evaluate the density matrix of the first order using the trial function (%
\ref{PMWF}) 
\begin{equation}
\rho =\frac{1}{L}\frac{\left\langle \psi \right| \sum\limits_{<ij>,\sigma
}\left( a_{i\sigma }^{\dagger }a_{j\sigma }+H.c.\right) \left| \psi
\right\rangle }{\langle \psi \left| \psi \right\rangle }.  \label{density}
\end{equation}

Here, there is a significant complication as compared to the Gutzwiller
trial wave function. While a fermion hops from site $i$ to site $j$, the
initial configurations of pair $i-j$ and adjacent pairs (e.g. $i-k$, $j-l$)
change (see Fig.1a). Let us fix a configuration of pair $i-j$ and adjacent
pairs $i-k$, $j-l$ and calculate function $W$ of residual lattice by means
of Eqs.(\ref{W}), (\ref{Q}), and (\ref{gamma}). The result is denoted by $%
W^{\prime }$. Then, a fraction of configurations containing the fixed
fragment can be written as follows 
\begin{equation}
\frac{W^{\prime }}{W}=y_{(ij)}\prod_{k}\left( \frac{y_{(ki)}}{x_{(i)}}%
\right) \prod_{j}\left( \frac{y_{(jl)}}{x_{(j)}}\right)  \label{WW}
\end{equation}
where $y_{(ij)}$ means some $y_{\lambda }$ corresponding to pair $(ij)$. The
term of the density matrix which comes from the transition from
configuration $1$ to configuration $2$ takes the form 
\begin{equation}
\frac{\prod_{i}g_{i}(2)}{\prod_{i}g_{i}(1)}\frac{W^{\prime }(1)}{W}
\label{gg}
\end{equation}
where the first factor is the ratio between amplitudes of configuration $1$
and $2$, i.e. $g_{i}(1)$ and $g_{i}(2)$ correspond to the configurations $1$
and $2$. In general, the procedure is similar to the Gutzwiller method \cite
{Gutzw} but the only parameter $g_{0}$ enters into Eq.(\ref{gg}) in the
latter case.

By means of Eqs. (\ref{WW}) and (\ref{gg}) one can sum up over all the
configurations and calculate the density matrix 
\begin{equation}
\rho =4\left[ 2y_{4}\left( a_{1}a_{2}\right) ^{z-1}+\frac{y_{3}g_{7}}{%
g_{0}g_{3}}a_{1}^{2(z-1)}+\frac{y_{7}g_{0}g_{3}}{g_{7}}a_{2}^{2(z-1)}\right]
,  \label{ro}
\end{equation}
where 
\[
a_{1}=\frac{y_{2}g_{4}+y_{3}g_{4}g_{3}^{-1}+y_{4}\left( g_{7}+1\right)
g_{4}^{-1}}{x} 
\]
and 
\[
a_{2}=\frac{y_{6}g_{4}+y_{7}g_{4}g_{7}^{-1}+y_{4}\left( g_{3}+1\right)
g_{4}^{-1}}{1/2-x}. 
\]

The first term of Eq.(\ref{ro}) describes transitions which do not change
the total number of doubly occupied sites. The second and third terms are
due to transitions corresponding to annihilation or creation of a doubly
occupied site. Let us exclude parameters $g_{i}$ by means of Eqs.(\ref{sys}%
). Then, after straightforward simplifications we obtain 
\begin{eqnarray}
\rho &=&8(y_{4}+\sqrt{y_{3}y_{7}})\left[ \frac{y_{4}}{x(1/2-x)}\right. 
\nonumber \\
&&\times \left. (\sqrt{y_{2}}+\sqrt{y_{3}}+\sqrt{y_{6}}+\sqrt{y_{7}})\right]
^{z-1}.  \label{roPM}
\end{eqnarray}

Finally it is convenient to present the total energy of Fermi system in
Gutzwiller's form 
\begin{equation}
E=\frac{1}{L}\frac{\left\langle \psi \left| H\right| \psi \right\rangle }{%
\left\langle \psi \mid \psi \right\rangle }=q\varepsilon _{0}+xU
\label{EnergyPM}
\end{equation}
where $q=\rho /\rho ^{0}$, $\rho ^{0}$ is the density matrix (\ref{roPM}) at 
$U=0$, 
\[
\varepsilon _{0}=2{\bf V}_{F}^{-1}\int_{{\bf V}_{F}}\varepsilon _{{\bf k}}%
{\normalsize d}{\bf k} 
\]
is the average energy of the non-interacting fermions. First we calculate $%
\rho ^{0}$ by minimization of the ground state energy at $U=0$, i.e. $\rho
^{0}=\min_{\left\{ x,y_{3},y_{4},y_{7}\right\} }\left( \rho \right) $. The
ground state energy at nonzero $U$ is determined as $\min_{\left\{
x,y_{3},y_{4},y_{7}\right\} }\left( E\right) $. The function $E$ turns out
to be smooth, without singular points within the domain of the function and
its minimum is easily found numerically. I used a refined Nelder-Mead
simplex algorithm for the minimum search.

The ground state energy of the PM phase calculated by this method is shown
in Fig.2: (a) a one-dimensional chain ($z=2$) with the dispersion law $%
\varepsilon _{{\bf k}}=-2\cos k_{x}$, (b) a square lattice ($z=4$), $%
\varepsilon _{{\bf k}}=-2[\cos k_{x}+\cos k_{y}]$, (c) a simple cubic
lattice ($z=6$), $\varepsilon _{{\bf k}}=-2[\cos k_{x}+\cos k_{y}+\cos
k_{z}] $. Symmetric and antisymmetric correlation functions of the nearest
neighbors 
\begin{eqnarray}
G_{s} &=&\left\langle n_{\uparrow }n_{\uparrow }\right\rangle ^{\prime
}+\left\langle n_{\downarrow }n_{\downarrow }\right\rangle ^{\prime
}=2\left( y_{2}+2y_{4}+y_{6}\right) ,  \nonumber \\
G_{a} &=&\left\langle n_{\uparrow }n_{\downarrow }\right\rangle ^{\prime
}+\left\langle n_{\downarrow }n_{\uparrow }\right\rangle ^{\prime }=2\left(
y_{2}+2y_{4}+y_{7}\right)  \label{corr}
\end{eqnarray}
are shown in Fig.3 for the same lattices as in Fig.2. The prime in Eqs.(\ref
{corr}) denotes the averaging over nearest-neighbor pairs only. Further
details of the calculations can be found in Ref.\cite{Kudasov}. This
technique was also applied to the AFM phase of the Hubbard model \cite
{Kudasov}.

\subsection{Low-energy spectrum and effective mass}

To investigate the energy spectrum of low-lying excitations in a model with
the total energy of the form (\ref{EnergyPM}), one can use the Fermi liquid
approach of Ref.\cite{Vollhardt}. Let $E_{g}$ be the ground state of the
system (\ref{EnergyPM}). Then, let us create a new initial state $|\varphi _{%
{\bf k}\sigma }\rangle =a_{{\bf p}\sigma }^{\dagger }a_{{\bf p}^{\prime
}\sigma }|\varphi _{0}\rangle $ and a new trial wave function 
\begin{equation}
\left| \psi _{{\bf k}\sigma }\right\rangle =\prod_{\lambda }g_{\lambda }^{%
\widehat{P}_{\lambda }}\left| \varphi _{{\bf k}\sigma }\right\rangle
\label{excited}
\end{equation}
where ${\bf k}={\bf p}-{\bf p}^{\prime }$, ${\bf p}$ and ${\bf p}^{\prime }$
are wave vectors lying above and below the Fermi surface, correspondingly. $%
|\varphi _{0}\rangle $ is the ground state of non-interacting fermions. It
should be noted that the new trial wave function has the same number of
fermions of each spin as the initial wave function (the canonical ensemble).
Since the operator on the right hand side of Eq.(\ref{new}) is
translationally invariant the new trial wave function has the well-defined
wave vector ${\bf k}$. Then, we perform the procedure developed above to
determine the minimum energy $E_{g}+\delta E_{{\bf k}\sigma }$ corresponding
to the excited state $\left| \psi _{{\bf k}\sigma }\right\rangle $, i.e. we
find a new stationary solution of the Hamiltonian (\ref{hubb}). It is easy
to see that $\delta E_{{\bf k}\sigma }$ is small, of the order of $1/N$. We
express the energy variation as 
\begin{eqnarray}
\delta E_{{\bf k}\sigma } &=&q\ \delta E_{{\bf k}\sigma }^{0}  \nonumber \\
&&+\varepsilon _{0}\left( \frac{\partial q}{\partial x}\delta x+\sum_{i}%
\frac{\partial q}{\partial \lambda _{i}}\delta \lambda _{i}\right) +U\
\delta x  \label{EnExp}
\end{eqnarray}
where $\lambda _{i}=y_{3},y_{4},y_{7}$ and $\delta E_{{\bf k}\sigma }^{0}$
is the energy variation of non-interacting fermions corresponding to the
excitation $|\varphi _{{\bf k}\sigma }\rangle $. Since the ground state is
minimal the following conditions are valid 
\begin{eqnarray}
\varepsilon _{0}\frac{\partial q}{\partial x}+U &=&0,  \nonumber \\
\frac{\partial q}{\partial \lambda _{i}} &=&0.  \label{EnEqu}
\end{eqnarray}

Combining Eqs.(\ref{EnExp}) and (\ref{EnEqu}) we find that $\delta E_{{\bf k}%
\sigma }=q\ \delta E_{{\bf k}\sigma }^{0}$ for any trial wave function
generated by Eq.(\ref{excited}), i.e. the low-lying energy spectrum of our
model is 
\[
\varepsilon _{{\bf k}\sigma }=q\ \varepsilon _{{\bf k}\sigma }^{0} 
\]
where $\varepsilon _{{\bf k}\sigma }^{0}$ is the energy spectrum of
non-interacting fermions. Retaining the terms of the second order
infinitesimal in the expansion (\ref{EnExp}) we obtain the effective Fermi
liquid theory \cite{Vollhardt}. One can perform the similar calculation with
the grand canonical ensemble ($|\varphi _{{\bf k}\sigma }\rangle =a_{{\bf k}%
\sigma }^{\dagger }|\varphi _{0}\rangle $) but it would be more complicated
because, in this case, a dependence of $q$ on the number of fermions has to
be taken into account. The effective mass at the Fermi surface is $m=q^{-1}$
where the effective mass of non-interacting fermions is assumed to be $%
m_{0}=1$. This result is in an agreement with the phenological Brinkman-Rice
approach \cite{Brinkman} and slave-boson treatment \cite{Kotliar}. The
effective mass in the half-filled Hubbard model is shown in Fig.4 as a
function of $U$. A detailed discussion of the excitation spectrum will be
presented elsewhere.

\section{Ground State of the Kondo-Hubbard lattice}

The Kondo-Hubbard lattice model can be expressed in the following form \cite
{Schork} 
\begin{eqnarray}
H_{{\normalsize KH}} &=&H_{{\normalsize H}}+J\sum_{i}{\bf S}_{i}^{l}{\bf S}%
_{i}^{c}  \nonumber \\
&=&H_{{\normalsize H}}+J\sum_{i}\left[ S_{iz}^{l}S_{iz}^{c}+\frac{1}{2}%
\left( S_{i+}^{l}S_{i-}^{c}+S_{i-}^{l}S_{i+}^{c}\right) \right] ,
\label{KHLM}
\end{eqnarray}
where $H_{\text{H}}$ is the Hubbard Hamiltonian, ${\bf S}_{i}^{c}$ is the
spin operator of an itinerant fermion at site $i$, ${\bf S}_{i}^{l}$ denotes
the spin or the total angular moment operator ($f$-spin) depending on the
nature of the localized state. We consider the PM phase at half
band-filling. It is convenient to represent the Kondo-Hubbard lattice as an
equivalent lattice in Fig.1b. Here, we obtain a new sort of nearest-neighbor
pairs, namely, itinerant fermion - localized fermion on the same site (open
circle - black circle in Fig.1b). It should be noted that there are no
additional closed paths in the lattice as compared to Fig.1a. That is why,
we don't bring about additional simplifications as compared to the Hubbard
model. The general form of the trial wave function of the Kondo-Hubbard
lattice can be presented as 
\begin{equation}
\left| \psi _{{\normalsize KH}}\right\rangle =g_{r}^{\widehat{\Re }%
_{r}}\left| \psi _{{\normalsize H}}\right\rangle \left| \varphi
_{l}\right\rangle =g_{r}^{\widehat{\Re }}\prod_{\lambda }g_{\lambda }^{%
\widehat{P}_{\lambda }}\left| \varphi _{c}\right\rangle \left| \varphi
_{l}\right\rangle  \label{KHLMWF}
\end{equation}
where $\left| \varphi _{c(l)}\right\rangle $ is the initial wave function of
itinerant ($c$) and localized fermions ($l$), $\widehat{P}_{\lambda }$ are
the projection operators for the Hubbard model (\ref{new}), $\left| \psi
_{H}\right\rangle $ is the trial wave function of the Hubbard model and $%
\widehat{\Re }=-4\sum_{i}S_{zi}^{l}S_{zi}^{c}$ is the new projection
operators for itinerant fermion - localized fermion pairs. Index $z$ denotes
the projection on $z$-axis. $\left| \varphi _{l}\right\rangle $ is the PM
phase without correlations (all spin configurations are equiprobable). In
the next two subsections, we shall define concretely the trial wave function
(\ref{KHLMWF}) for two cases.

\subsection{$S=1/2$ Kondo-Hubbard lattice}

There are three eigenstates of operator ${\bf S}_{i}^{l}{\bf S}_{i}^{c}$ in
case of the spin $1/2$ localized state (singlet, triplet and $S_{iz}^{c}=0,$
see Table \ref{Table2}). Let us introduce the eigenvalues of the operator $%
r_{0}$, $r_{1}$ and $r_{2}$. The self-consistency and normalization
conditions (\ref{self}), (\ref{norm}) remain valid and new ones appear 
\begin{equation}
r_{0}=x,\text{ }r_{1}+r_{2}=1/2-x.  \label{self2}
\end{equation}

From this it follows that a new independent parameter $r=r_{1}$ appears in
addition to the set describing the Hubbard subsystem ($x,$ $y_{3},$ $y_{4},$ 
$y_{7}$). Then, the trial wave function takes the form 
\begin{eqnarray}
\left| \psi _{{\normalsize KH}}\right\rangle &=&g_{r}^{\widehat{\Re }}\left|
\psi _{{\normalsize H}}\right\rangle \left| \varphi _{l}\right\rangle 
\nonumber \\
&=&g_{r}^{\widehat{\Re }}g_{0}^{\widehat{X}}g_{3}^{\beta _{3}\widehat{Y}%
_{3}}g_{4}^{\beta _{4}\widehat{Y}_{4}}g_{7}^{\beta _{7}\widehat{Y}%
_{7}}\left| \varphi _{c}\right\rangle \left| \varphi _{l}\right\rangle .
\label{KHLMWF1/2}
\end{eqnarray}
and its norm is 
\begin{eqnarray}
\left\langle \psi |\psi \right\rangle &=&\sum_{\left\{
r,x,y_{3},y_{4},y_{7}\right\} }W_{\left\{ r,x\right\} }^{K}W_{\left\{
x,y_{3},y_{4},y_{7}\right\} }^{H}  \nonumber \\
&&\times
g_{r}^{4Lr}g_{0}^{2Lx}g_{3}^{2zLy_{3}}g_{4}^{8zLy_{4}}g_{7}^{2zLy_{7}} 
\nonumber \\
&=&\sum_{\left\{ r,x,y_{3},y_{4},y_{7}\right\} }R_{\left\{
r,x,y_{3},y_{4},y_{7}\right\} }  \label{normKHL}
\end{eqnarray}
where $W^{K(H)}$ is the Kondo (Hubbard) weight of the set. $W^{H}$ is from
Eqs.(\ref{Q}) and (\ref{gamma}). The Kondo weight can be easily calculated 
\begin{equation}
W^{K}=\frac{\left( 2xL\right) !}{\left[ \left( xL\right) !\right] ^{2}}%
\left\{ \frac{\left[ L\left( 1/2-x\right) \right] !}{\left[ rL\right]
!\left[ L\left( 1/2-x-r\right) \right] !}\right\} ^{2}.  \label{WKHL}
\end{equation}

Using Eqs. (\ref{normKHL}) and (\ref{WKHL}) we search the global maximum of
the norm following the procedure described in the previous section. We
obtain the system of equations (\ref{dlnR}) where $\eta _{\lambda
}=r,x,y_{3},y_{4},y_{7}$. The equations for parameters $g_{3},$ $g_{4}$ and $%
g_{7}$ remain the same as in system (\ref{sys}). For the other parameters we
obtain 
\begin{eqnarray}
g_{0} &=&\frac{1/2-x}{2\sqrt{r\left( 1/2-x-r\right) }}  \nonumber \\
&&\times \left( \frac{1/2-x}{x}\right) ^{z-1}\left( \frac{x-y_{3}-2y_{4}}{%
1/2-x-y_{7}-2y_{4}}\right) ^{z/2},  \nonumber
\end{eqnarray}
\begin{equation}
g_{r}^{4}=\frac{r}{1/2-x-r}.  \label{sys12}
\end{equation}

The total energy of the Kondo-Hubbard lattice includes Hubbard and exchange
parts 
\begin{equation}
E=\frac{1}{L}\frac{\left\langle \psi \left| H\right| \psi \right\rangle }{%
\left\langle \psi \mid \psi \right\rangle }=q\varepsilon _{0}+xU+J(\rho
_{zz}+\rho _{\pm })  \label{EKHL}
\end{equation}
where $q=\rho _{H}/\rho _{H}^{0}$, $\rho _{H}^{0}$ is the Hubbard density
matrix at $U=0$, $\rho _{zz},$ $\rho _{\pm }$ are the density matrices
corresponding to $zz$ and spin-flip interactions. Since there are new bonds
in the lattice in Fig.1b, the Hubbard density matrix is different from that
of the Hubbard model. Thus, instead of Eq.(\ref{ro}) we get 
\begin{eqnarray}
\rho _{H} &=&4\left[ 2y_{4}\left( a_{1}a_{2}\right) ^{z-1}b_{1}b_{2}\right. 
\nonumber \\
&&\left. +\frac{y_{3}g_{7}}{g_{0}g_{3}}a_{1}^{2(z-1)}b_{2}^{2}+\frac{%
y_{7}g_{0}g_{3}}{g_{7}}a_{2}^{2(z-1)}b_{1}^{2}\right]  \label{roHKHL}
\end{eqnarray}
where 
\begin{eqnarray*}
b_{1} &=&rg_{r}^{-1}+g_{r}\left( 1/2-x-r\right) , \\
b_{2} &=&\frac{1}{2}\left( g_{r}+g_{r}^{-1}\right) .
\end{eqnarray*}
Straightforward calculations of exchange terms give 
\begin{eqnarray}
\rho _{zz} &=&-\frac{1}{2}\left( 2r+x-1/2\right) ,  \label{roHzzpp} \\
\rho _{\pm } &=&-\frac{2y_{4}+y_{6}g_{7}+y_{7}g_{7}^{-1}}{\left(
1/2-x\right) ^{z}}.  \nonumber
\end{eqnarray}

Expressions (\ref{EKHL}), (\ref{roHKHL}) and (\ref{roHzzpp}) present the
total energy of the Kondo-Hubbard lattice in an analytic form as a function
of independent variational parameters $r,$ $x,$ $y_{3},$ $y_{4},$ $y_{7}$.
The ground state energy is the global minimum of Eq.(\ref{EKHL}) with
respect to these parameters. The spin nearest-neighbor correlation function
of the itinerant subsystem 
\begin{equation}
G_{c}=\left\langle S_{zi}^{c}S_{zj}^{c}\right\rangle ^{\prime }=\left(
G_{s}-G_{a}\right) /4=\frac{1}{2}\left( y_{6}-y_{7}\right)  \label{Gc}
\end{equation}
and the spin correlation function of localized nearest neighbors are shown
in Fig.5. The last is calculated by means of the superposition hypothesis 
\cite{Ziman} 
\begin{eqnarray}
G_{l} &=&\left\langle S_{zi}^{c}S_{zj}^{c}\right\rangle ^{\prime }\left[
\left\langle S_{zi}^{c}S_{zi}^{l}\right\rangle \right] ^{2}  \nonumber \\
&=&\frac{1}{2}\left( 1-2x-4r\right) ^{2}\left( y_{6}-y_{7}\right) .
\label{Gl}
\end{eqnarray}

Following the expansion (\ref{EnExp}) we determine the effective mass of the
itinerant fermions as $m=q^{-1}$. It is plotted against $J$ in Fig.6. We
have also plotted the fraction of lattice sites occurred to be in the Kondo
singlet state ($2r$) in Fig.7.

\subsection{$S=5/2$ Kondo-Hubbard lattice}

Bearing in mind cerium Kondo-Hubbard lattices (Ce$^{3+}$) we generalize the
technique developed above to spin 5/2 (the total angular moment or $f$%
-spin). The crystal field is neglected in the Hamiltonian (\ref{KHLM}).
Crystal field effects will be discussed briefly in the next section. Let us
compile a table of all intrasite configurations (see Table \ref{table3}).
Since the Hamiltonian (\ref{KHLM}) is rotation invariant, possibilities of
configurations with $S_{iz}^{c}=0$ are equal each other, $r_{0}=x/3$. There
are 6 configurations with $S_{iz}^{c}\neq 0$. They are bound by the
normalization condition 
\begin{equation}
\sum_{i=1}^{6}r_{i}=1/2-x.  \label{normKHL52}
\end{equation}

It follows that one of $r_{i}$ is dependent (we take $r_{6}$ as the
dependent parameter). The general trial wave function (\ref{KHLMWF1/2})
remain valid for the spin 5/2. Then, its norm is 
\begin{eqnarray}
\left\langle \psi |\psi \right\rangle =\sum_{\left\{
r_{i},x,y_{3},y_{4},y_{7}\right\} }W_{\left\{ r_{i},x\right\}
}^{K}W_{\left\{ x,y_{3},y_{4},y_{7}\right\}}^{H}  \nonumber \\
\times g_{r}^{4L\sum_{i=1}^{6}r_{i}\left( 7-2i\right)
}g_{0}^{2Lx}g_{3}^{2zLy_{3}}g_{4}^{8zLy_{4}}g_{7}^{2zLy_{7}}  \label{nmr52}
\end{eqnarray}

In case of $S=5/2$, the Kondo part of the configuration weight is reduced to 
\begin{equation}
W^{K}=\frac{\left( 2xL\right) !}{\left[ \left( xL\right) !\right] ^{6}}%
\left\{ \frac{\left[ L\left( 1/2-x\right) \right] !}{\prod_{i}\left(
r_{i}L\right) !}\right\} ^{2}.  \label{W52}
\end{equation}

The necessary condition of the global maximum of $R$ (\ref{dlnR}), where $%
\eta _{\lambda }=r_{i},x,y_{3},y_{4},y_{7}$, leads to a system of nine
equations. In particular, there are five equations for independent $r_{i}$%
\begin{equation}
g_{r}^{2\left( 12-2i\right) }=\frac{r_{i}}{r_{6}}  \label{gr52}
\end{equation}
where $r_{6}=1/2-x-\sum_{i}r_{i}$. From here one can see that, in effect,
there is the only independent parameter $r$. Let it be $r=r_{1}$. This is no
wonder because the only variational parameter $r$ enters into the trial wave
function. The five equations for $r_{i}$ gives the nonlinear equation
relating $r$ and $g_{r}$%
\begin{equation}
r\left( 1+g_{r}^{-4}+g_{r}^{-8}+g_{r}^{-12}+g_{r}^{-16}+g_{r}^{-20}\right)
=1/2-x,  \label{nonlin}
\end{equation}
The equation for $g_{0}$ transforms to 
\begin{eqnarray}
g_{0} &=&\frac{1/2-x}{6\sqrt{rr_{6}}}\left( \frac{1/2-x}{x}\right) ^{z-1} 
\nonumber \\
&&\times \left( \frac{x-y_{3}-2y_{4}}{1/2-x-y_{7}-2y_{4}}\right) ^{z/2}.
\label{sys52}
\end{eqnarray}
The equations for $y_{3}$, $y_{4}$, $y_{7}$ are the same as for the Hubbard
model (\ref{sys}).

The density matrix falls into the Hubbard, $zz$ and spin-flip terms in the
same manner as for the localized spin $1/2$. The general form of the Hubbard
term is Eq.(\ref{roHKHL}) where 
\begin{eqnarray*}
b_{1} &=&\frac{1}{6}%
\mathop{\displaystyle \sum }%
\limits_{i=1}^{3}\left( g_{r}^{2i-1}+g_{r}^{1-2i}\right) , \\
b_{2} &=&\frac{r}{1/2-x}%
\mathop{\displaystyle \sum }%
\limits_{i=1}^{6}g_{r}^{-3-2i}.
\end{eqnarray*}

Straightforward calculations give the $zz$ and spin-flip terms 
\begin{eqnarray}
\rho _{zz} &=&-\frac{1}{2}\sum_{i=1}^{6}rg_{r}^{4(1-i)}(7-2i),  \label{ro52}
\\
\rho _{\pm } &=&-(1-2x-rg_{r}^{-20})g_{r}^{-2}.  \nonumber
\end{eqnarray}

The total energy is written in the form of Eq.(\ref{EKHL}). The ground state
energy is determined by the minimization over the independent variational
parameters. There is a complication of the numerical procedure because in
the present case the variational parameter $g_{r}$ can't be expressed
analytically in terms of $r$. In other respects the calculations are similar
to that discussed in previous sections. The spin correlation function of
itinerant fermions (\ref{Gc}) and that of localized states 
\begin{eqnarray}
G_{l} &=&\left\langle S_{zi}^{c}S_{zj}^{c}\right\rangle ^{\prime }\left[
\left\langle S_{zi}^{c}S_{zi}^{l}\right\rangle \right] ^{2}  \nonumber \\
&=&\frac{1}{4}\left( y_{6}-y_{4}\right) r^{2}  \nonumber \\
&&\times \left( 5\left[ 1-g_{r}^{-20}\right] +3\left[
g_{r}^{-4}-g_{r}^{-16}\right] +g_{r}^{-8}-g_{r}^{-12}\right) ^{2}
\label{Gl52}
\end{eqnarray}
are shown in the insert in Fig.5. The effective mass and the function $2r$
are plotted in the inserts in Fig.6 and Fig.7.

\section{DISCUSSION AND CONCLUSIONS}

The trial wave functions used in the present approach (\ref{PMWF}), (\ref
{KHLMWF}) have a remarkable property. The operator on the right side of Eq.(%
\ref{new}) commutes with the operators of the crystal point group
(rotations, reflections, inversion) and the translation group. That is why
the trial wave function retains all the symmetries of the initial wave
function. In contrast to the Gutzwiller wave function, we have incorporated
the nearest-neighbor correlations into the trail wave function. That is why,
a local structure surrounding an atom appears. A similar trial wave function
was used in Ref.\cite{Fulde1,Fulde}. The ground state energy was evaluated
there by means of an expansion of exponent $\exp (-\gamma \widehat{Y}%
_{\lambda })$. To perform this a systematic diagram representation was
introduced \cite{Fulde1,Fulde}. This approach turned out to be very
successful in quantum chemistry for small molecules because one obtains an
excellent results taking into account the first terms of the series. At the
same time, the eigenvalues of $\widehat{Y}_{\lambda }$ are approximately
proportional to the number of lattice sites. That is why, it is hardly
possible to use this approach in the thermodynamic limit ($L\rightarrow
\infty $) for the strong short-range order. Let us consider a set of
approximations in the framework of Kikuchi's method, namely the Gutzwiller
approximation (a cluster consists of the only site) as the first one, a pair
Kikuchi's approximation (a cluster consists of two sites) as the second one,
and etc. On each step, a cluster becomes larger. It was rigorously proved
for 2D square lattice and 3D cubic lattice that in the thermodynamic limit
we approach the rigorous solution while a cluster size increases \cite{kik2}%
. This allows us to go beyond the infinite dimension limit.

We calculated the ground state energy of the Hubbard model in the PM phase
for a one-dimensional chain, square and simple cubic lattices (see Fig.2).
The result for a one-dimensional chain is very close to that obtained by an
analytic investigation of Gutzwiller wave function \cite{metz2}. Let us
mention that the method presented above gives much lower the ground state
energy in the AFM phase \cite{Kudasov}. The results for the square and
simple cubic lattices are compared with that of the VMC method \cite{Yokoy}
in Fig.2b,c. Since the VMC method is based on the Gutzwiller trial wave
function the difference between this approach and our method is due to
effects of the short-range correlations. It can be seen that near $%
U_{C}=8\left| \varepsilon _{0}\right| $ (i.e. the critical value of $U$ in
the Gutzwiller approximation) the ground state energy of the trial wave
function Eq.(\ref{PMWF}) is substantially lower (two or tree times) than
that obtained by the VMC method, i.e. the short-range order considerably
reduces the ground state energy. This result is in contrast to the $1/D$%
-expansion in the dynamical mean field theory \cite{Gebh} and shows that the
perturbation theory methods are hardly suitable here.

The narrow quasi-particle band (see Fig.4) is appeared at the Fermi level
similarly to the results of Gutzwiller's approach \cite{Gutzw,Vollhardt}. At
large $U$, the effective mass become linear on $U$. The slope of the
function $m(U)$\ increases while the lattice dimension becomes larger. The
limit of infinite dimensions of the ground state (\ref{PMWF}) was
investigated in Ref.\cite{Kudasov}. While the lattice dimension increases,
the ground state energy approaches to Gutzwiller's solution. We also note
that an exchange hole exists in the variational solution even at $U=0$ ($%
G_{s}<0.5$). It increases monotonically while $U$ grows
(exchange-correlation hole). Unlike the well-known Hubbard III solution the
short-range AFM correlations do not disappear in the limit $\left| t\right|
/U\ll 1$ but tend to a certain constant value (see Fig.3). In this limit,
the Hubbard model at half band-filling reduces to the spin-$1/2$ Hiesenberg
model. The residual AFM correlations in the PM phase are consistent with the
results of the ground state studying of the Hiesenberg model \cite{hies}. It
was also shown in Ref.\cite{Kudasov} that in the AFM phase, the both methods
(the VMC method and the present one) give almost the same ground state
energy, i.e. in the presence of the long-range order, the short-range
correlations are inessential.

In the Kondo-Hubbard model we observe two different types of behavior
depending on value of $J$. At small $J$, the exchange term weakly affects
the ground state energy and effective mass. It should be noted that the AFM
correlations between localized states at a pair of nearest-neighbors sites ($%
G_{l}$) increase with increasing of $J$. This effect appears because the
short-range AFM correlations between band fermions exist at $J=0$ and at
small $J$ the correlations between localized states follow the correlations
of band fermions. The growth of the exchange leads to suppression of the
band correlations. In Fig.5, it is seen a transition to another regime
(large $J$). For simple cubic lattice we have obtained a discontinuity of $%
G_{c}$ and a sharp turn of $G_{l}$ for both $S^{l}=1/2$ and $S^{l}=5/2$
cases, i.e. a phase transition. A smooth crossover is observed in 1D chain
and square lattice at this point (see Fig.5). At present, it is hardly
possible to establish a kind of the phase transition. We can rule out the
first order transition only because the ground state energy is smooth (the
derivative of the ground state energy is continuous). At higher $J$, the
correlations between localized states start to decrease with increasing $J$.
It should be mentioned that the behavior of the spin correlation function $%
G_{l}$ is similar to that ensued from Doniach's phase diagram despite the
fact that the RKKY interaction is not included into the variational theory.

The Coulomb interaction between itinerant fermions $U$ influences the ground
state in two ways. On the first hand, it reinforces the short-range order
and increases the AFM correlations between nearest-neighbor localized
states. On the other hand, at large $U$ the effective quasi-particle band
gets smaller that favors the Kondo regime.

At large $J$, the ground state energy approaches asymptotically the energy
of the pair singlet state ($\frac{3}{4}J$ for $S^{l}=1/2$). The probability
of the Kondo singlet ($2r$) at a site goes to 1 (see Fig.7) for $S^{l}=1/2$
and to some fixed value (independent on the lattice dimension) for $%
S^{l}=5/2 $. The difference is due to the fact that, in the first case, the
ground state is a superposition of two antiparallel states ($\frac{1}{\sqrt{2%
}}(\left| \uparrow \downarrow \right\rangle -\left| \downarrow \uparrow
\right\rangle )$) and, in the case of a pair $S^{l}=5/2$, $S^{c}=1/2$, it
involves all the possible states ($\left| S_{zi}^{l}=\frac{5}{2}%
\right\rangle \left| S_{zi}^{c}=-\frac{1}{2}\right\rangle $, $\left|
S_{zi}^{l}=\frac{3}{2}\right\rangle \left| S_{zi}^{c}=-\frac{1}{2}%
\right\rangle $ and {\it etc.}) for the total energy to be minimal. At large 
$J$ the effective mass of itinerant fermions increases drastically and tends
to almost linear behavior with variation $J$ ($m\varpropto \alpha J$ where $%
\alpha $ is the constant depending on the type of lattice) as can be seen in
Fig.6. This is the Kondo lattice regime. In the case of $S^{l}=5/2$, the
effective mass is significantly larger then for $S^{l}=1/2$. In a real
system, $m$ can't increase infinitely and a localization of fermions should
occur due to disorder or temperature effects but this is beyond the scope of
the present paper.

The crystal field can be taken into account in the variational scheme. It
can be easily done in the limit of the strong crystal field when we consider
the lowest level of the multiplet only. In the general case, we have to use
additional variational parameters to control different populations at
multiplet levels. For instance, if the ground state of Ce$^{3+}$ ion is
splitted by the crystal field into a doublet and a quadruplet (e.g. $\Gamma
_{7}$ and $\Gamma _{8}$ states), we should apply the only additional
variational parameter to the trail wave function. If it is splitted into
three doublets, we have to use two additional parameters and {\it etc}.

In the framework of the variational theory presented above the strongly
correlated coherent metal state appears in a natural way similarly to the
Gutzwiller theory. This provides a good basis for investigations of strongly
correlated metal systems and dense Kondo systems by means of this theory.
The main shortcoming of our approach is neglect of closed loops while we
treat the correlations on a lattice. Nevertheless, it is considered that
Kikuchi's method yields a good approximation if the correlation length is
not greater than a cluster size \cite{shlijper}. The well-developed cluster
variation method allows to include short closed loops, which are most
important, into considerations \cite{kik3}. The short-range correlations are
observed directly by the neutron scattering. The AFM correlations in the
strongly correlated metals like V$_{2}$O$_{3}$ and Kondo lattices turn out
to be strong but very short \cite{Bao,Aeppli,Aeppli2,Regn,Bernh}. That is
why the closed loops are inessential in this substances. Let us mention that
one can calculated the neutron cross-section and dynamical susceptibility
from the results obtained above. At the same time, it is hardly possible to
apply the present theory to systems with lengthy AFM correlations (e.g.
strongly underdoped high temperature superconductors \cite{HTSC}).

\section{Acknowledgments}

I is grateful to Prof. J. Brooks and Dr. W. Lewis for an invaluable support.
I wish to thanks Prof. G. Uimin for very useful discussions of the magnetism
of rare-earth compounds and Prof. P.Fulde for hospitality during the stay at
the Max-Plank-Institut f\"{u}r Komplexer Systeme (Dresden).

This work was carried out under the Project \#829 of the International
Science and Technology Center.

\section{FIGURE CAPTIONS}

Fig.1. (a) A fragment of $z=4$ lattice, (b) representation of $z=4$
Kondo-lattice. Itinerant and localized states are shown as light and black
circles, correspondingly; double lines denote the exchange interaction.

Fig.2. The ground state energy of the Hubbard model at half-band filling.
(a) A one-dimensional chain: the Gutzwiller solution (1), present study (2),
the exact solution; (b) square and (c) simple cubic lattices: the Gutzwiller
solution (1), the VMC method (2), the present study (3).

Fig.3. The symmetric $G_{s}$ (dashed lines) and antisymmetric $G_{a}$ (solid
lines) correlation functions of the Hubbard model for a one-dimensional
chain (1), square (2) and simple cubic lattices (3).

Fig.4. The effective mass in the half-filled Hubbard model: a
one-dimensional chain (1), square (2) and simple cubic lattices (3). The
dashed lines are guides to eye.

Fig.5. The spin correlation functions $G_{c}$ (dashed lines) and $G_{l}$
(solid lines) for the half-filed $S=1/2$ Kondo-Hubbard model: (a)\_a
one-dimensional chain, (b) square and (c) simple cubic lattices; $U=0.5U_{c}$%
. The correlation functions for the half-filed $S=5/2$ Kondo-Hubbard model
are shown in the insert.

Fig.6. The effective mass in the $S=1/2$ Kondo-Hubbard model: a
one-dimensional chain( 1), square (2) and simple cubic lattices (3); $%
U=0.5U_{c}$. The dashed lines are guides to eye. The effective mass for the
half-filed $S=5/2$ Kondo-Hubbard model is shown in the insert.

Fig.7. The probability of the Kondo singlet state in the $S=1/2$
Kondo-Hubbard model ($2r$). The probability of $\left| S_{zi}^{l}=\pm
5/2\right\rangle \left| S_{zi}^{c}=\mp 1/2\right\rangle $ states in the $%
S=5/2$ Kondo-Hubbard model is shown in the insert.

\begin{table}[tbp]
\caption{Pair projection operators, corresponding configurations and the
degeneracy factor}
\label{table}%
\begin{tabular}{dddddd}
Operator  &\multicolumn{2}{c}{Configuration}& Degeneracy \\
$\widehat{Y}_{i}$ &Site A&Site B& $\beta_{i}$\\
\tableline
$\widehat{Y}_{1}$ & $-$ & $-$ &1 \\
$\widehat{Y}_{2}$& $\uparrow$ $\downarrow$& $\uparrow$ $\downarrow$ &1\\
$\widehat{Y}_{3}$& $\uparrow$ $\downarrow$& $-$ &2\\
$\widehat{Y}_{4}$& $\uparrow$ & $-$ &2\\
$\widehat{Y}_{5}$& $\downarrow$ & $-$ &2\\
$\widehat{Y}_{6}$& $\uparrow$ &$\uparrow$ &1\\
$\widehat{Y}_{7}$& $\uparrow$ &$\downarrow$ &2\\
$\widehat{Y}_{8}$& $\uparrow$ $\downarrow$ &$\uparrow$ &2\\
$\widehat{Y}_{9}$& $\uparrow$ $\downarrow$ &$\downarrow$&2\\
$\widehat{Y}_{10}$& $\downarrow$& $\downarrow$ &1\\
\end{tabular}
\end{table}

\begin{table}[tbp]
\caption{Intrasite configurations of the spin $1/2$ Kondo-Hubbard lattice}
\label{Table2}%
\begin{tabular}{dddddd}
\multicolumn{2}{c}{Configuration}& Eigenvalue  & Degeneracy \\
Itinerant state&Localized state $S_{z}$&   & $ \beta_{i}$\\
\tableline
 $S=0$   & $\uparrow$ &$x$&2 \\
 $\uparrow$ & $\uparrow$ &$1/2-r-x$&2\\
 $\uparrow$ & $\downarrow$ &r&2\\
\end{tabular}
\end{table}

\begin{table}[tbp]
\caption{Intrasite configurations of the spin $5/2$ Kondo-Hubbard lattice}
\label{table3}%
\begin{tabular}{dddddd}
\multicolumn{2}{c}{Configuration}& Eigenvalue  & Degeneracy \\
Itinerant state $S^{c}_{z}$&Localized state $J^{l}_{z}$&   & $ \beta_{i}$\\
\tableline
 $0$   & any state &$x$&2 \\
 $1/2$ & $-5/2$ &$r_1$&2\\
 $1/2$ & $-3/2$ &$r_2$&2\\
$1/2$ & $-1/2$ &$r_3$&2\\
 $1/2$ & $1/2$ &$r_4$&2\\
$1/2$ & $3/2$ &$r_5$&2\\
 $1/2$ & $5/2$ &$r_6=1/2-x$ & 2 \\
 $ $      &  $$      &$-\sum_{i=1}^{5}r_i$ & \\

\end{tabular}
\end{table}

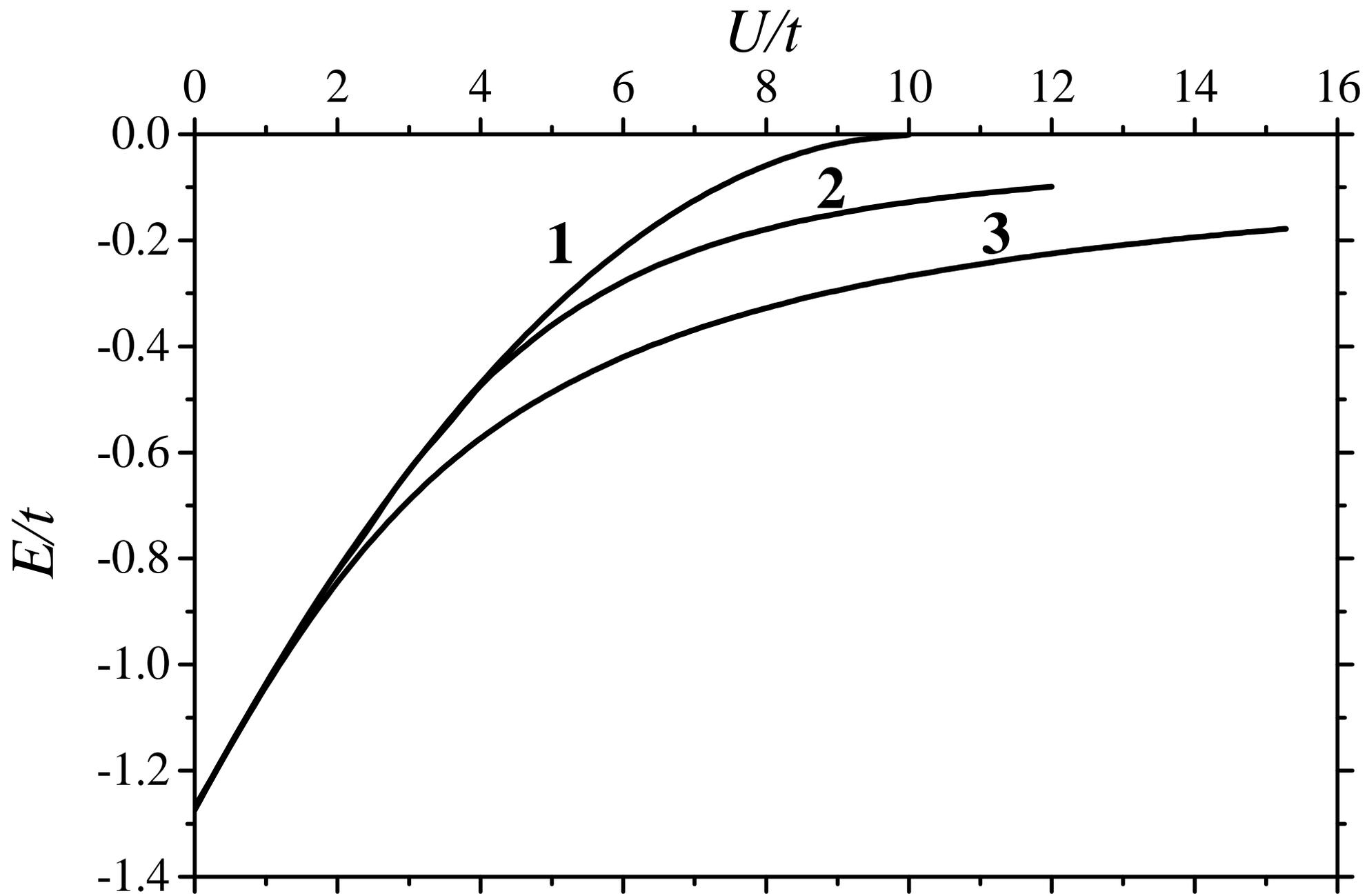
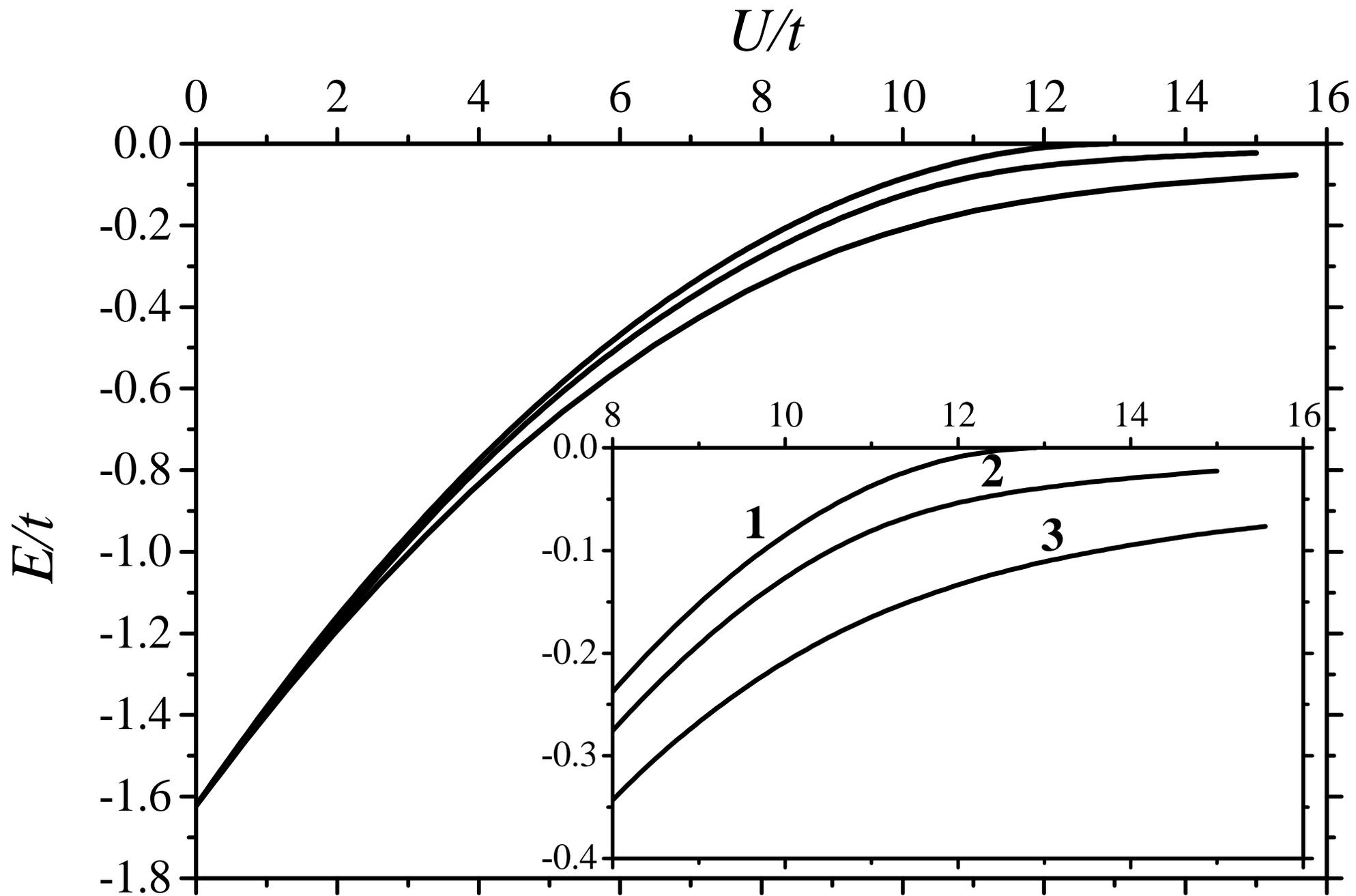
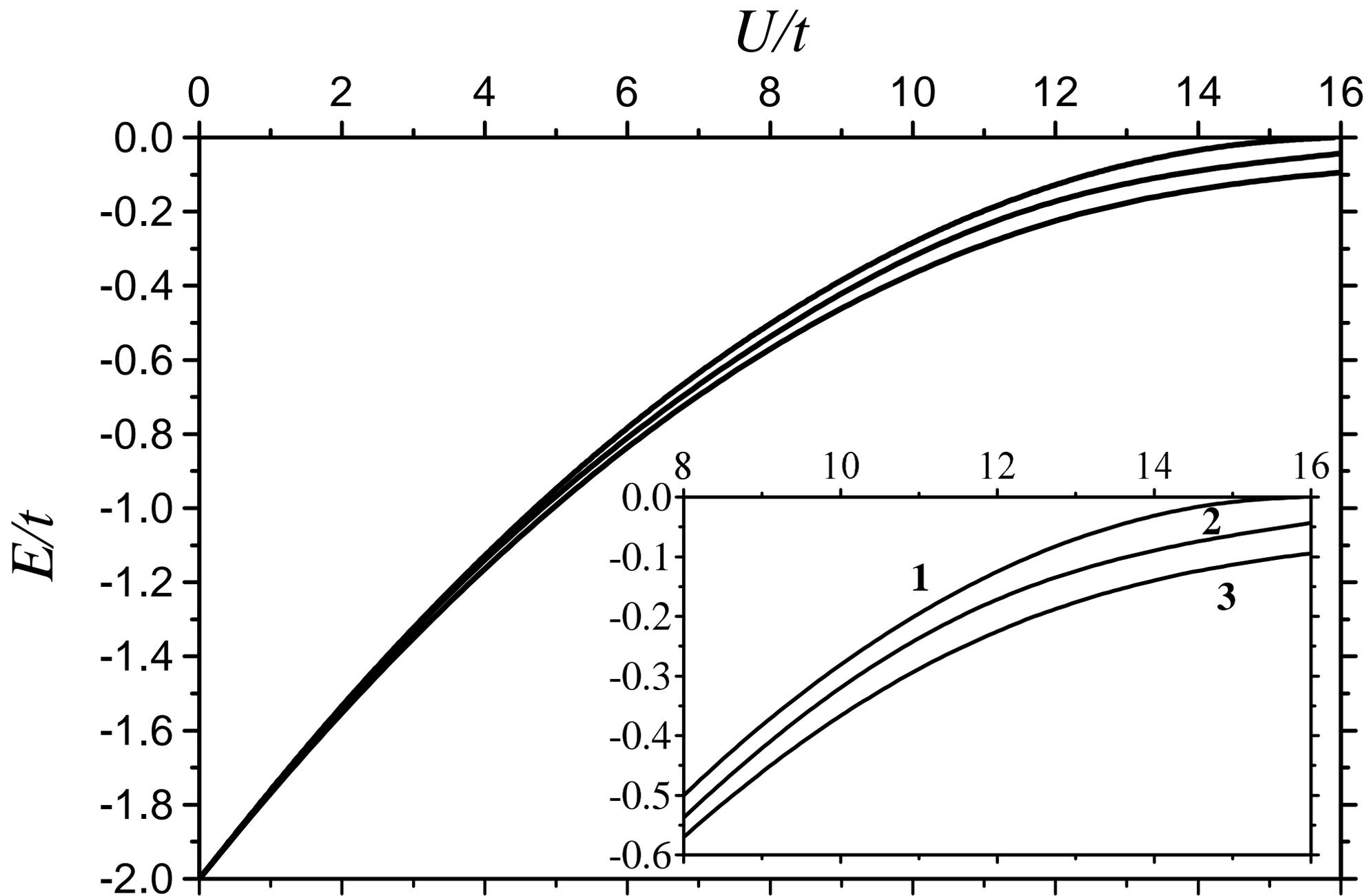
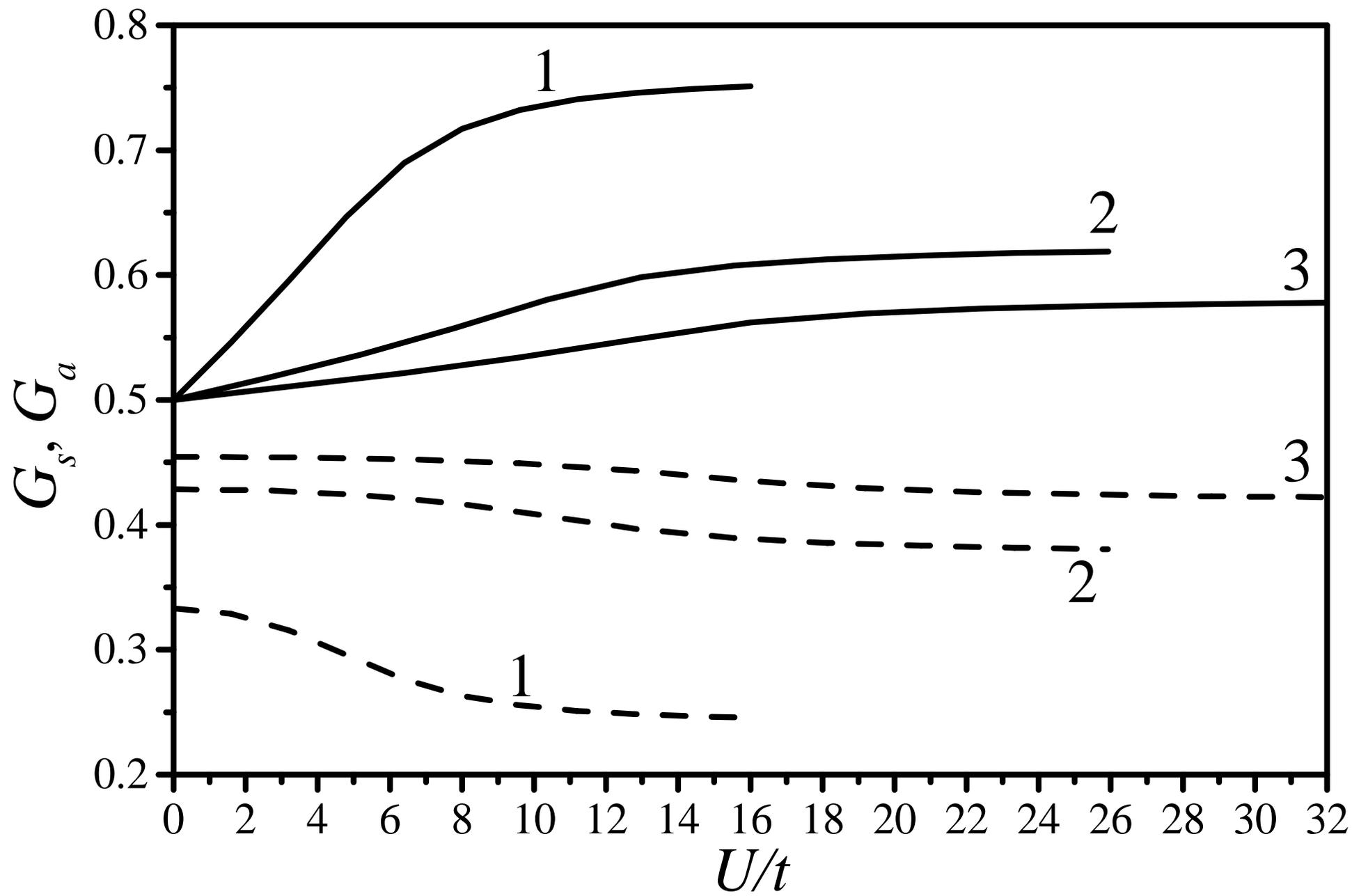
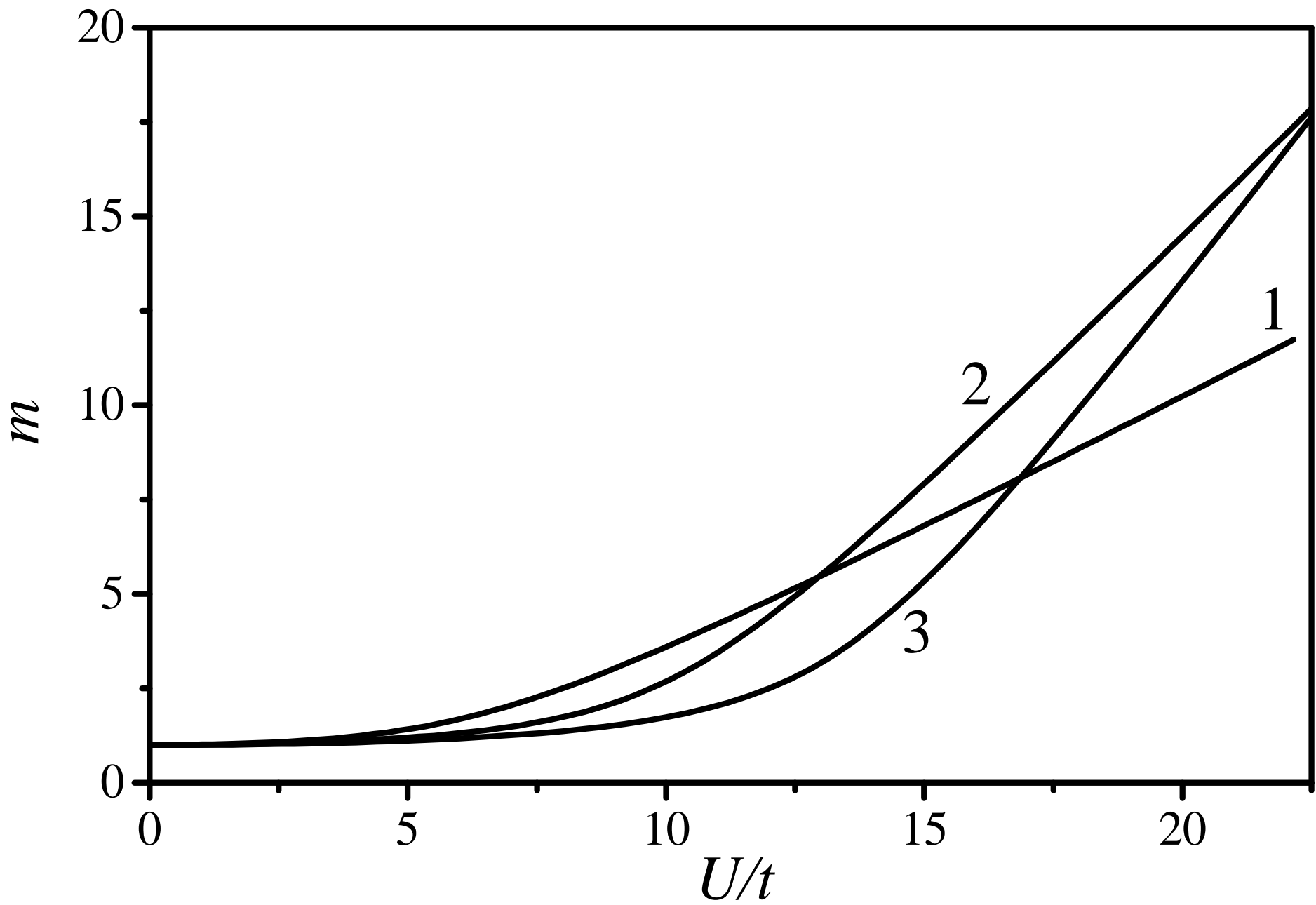
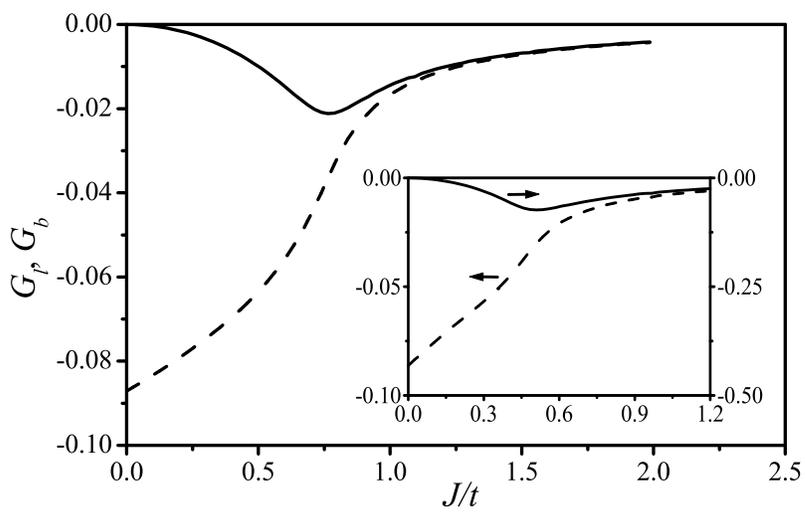
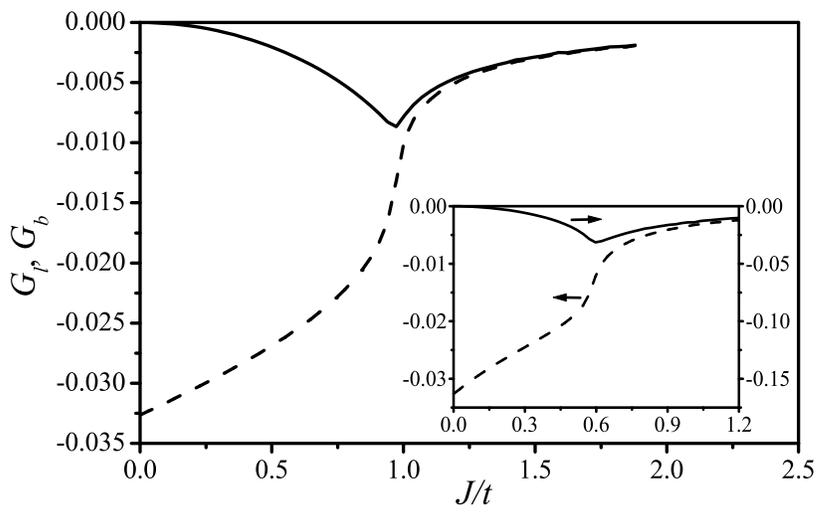
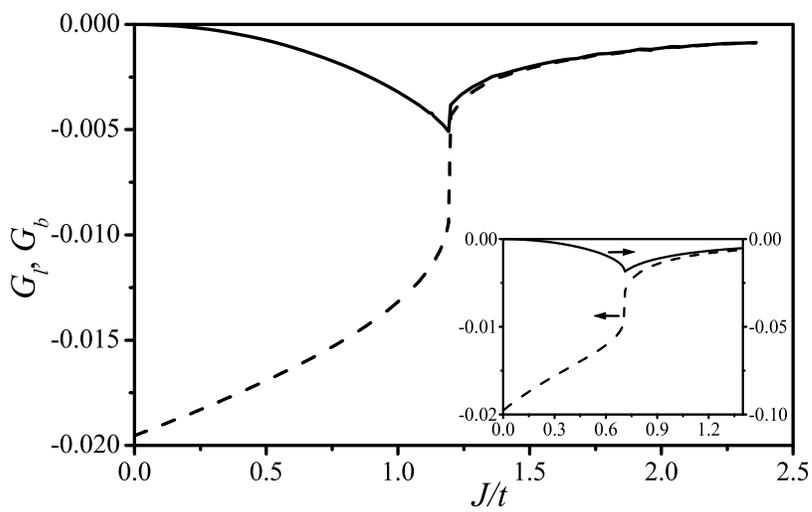
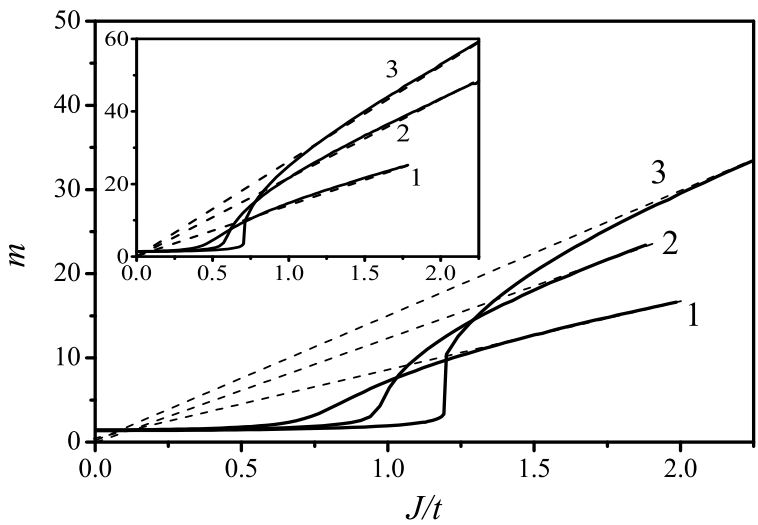
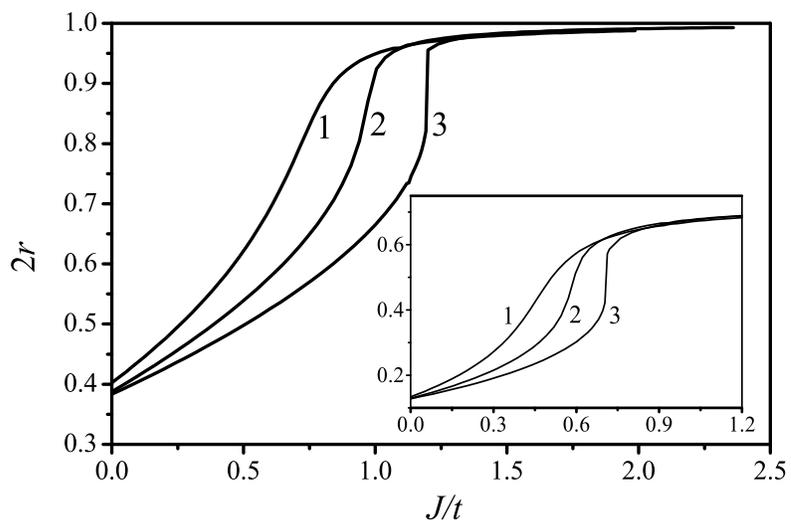
\end{document}